\documentclass[final,12pt,letterpaper]{JHEP}
\usepackage{amsmath,amssymb}
\usepackage[final]{epsfig}

\allowdisplaybreaks[1]
\newcommand\skipthis[1]{{}}

\chardef\til=`~

\newcommand{\eq}{\begin{equation}}
\newcommand{\en}{\end{equation}}
\newcommand{\eqa}{\begin{eqnarray}}
\newcommand{\ena}{\end{eqnarray}}
\newcommand{\eqs}{\begin{displaymath}}
\newcommand{\ens}{\end{displaymath}}
\newcommand{\eqas}{\begin{eqnarray*}}
\newcommand{\enas}{\end{eqnarray*}}

\vspace{-3cm}

\vspace{1cm}

\title{A Note on the Path Integral Representation \\
  of the Boundary State of D-brane}
\author{Yong Zhang\thanks{\tt zhangyo@itp.ac.cn
 }\\
  Institute of Theoretical Physics,  Academia Sinica, \\
 PO Box 2735,  Beijing
100080,  P. R.  China}

\abstract{
In this paper we construct path integral representations
of the boundary states in some special backgrounds such as the $U(1)$
gauge field background, the linear dilaton background and the open string tachyon 
background. The initial purpose of this paper is to construct a general solution of the 
boundary conformal field theory with the analytical approach, mainly for the 
constraint equations$(L_{n}-\tilde{L}_{-n}) |B \rangle =0 $
are difficult to be solved to obtain the solution represented by  string 
modes from the pure algebraic approach. However in the path integral representation 
it is easy transforming those algebraic
equations into the differential equations which can be solved.
Another purpose of this paper is to try to explore an open question.
we do not know how to construct an exact theory
of D-branes in the general background  until now. However
many recent researches show the 
boundary state description indeed seizes some fundamental features 
of D-branes in the rather special backgrounds. Since
the general background field effects can be easily introduced in  the path integral
representation, we argue that path integral representation of  the boundary
state should provide an exact description of D branes in  the general backgrounds.}

\preprint{hep-th/0005176}

\begin{document}

\section{Introduction}

It is well known
that D-branes in\cite{pol} play very important roles in the study of nonperturbative 
string theory 
dynamics.  However, we do not know how to construct an exact theory
of D-brane in the general background  until now. So  we have to limit ourselves 
within a very small circle, namely in some special backgrounds. This kind of study
may have some heuristic 
implications in seeking a good theory describing 
 D-brane. On the other hand
many recent researches show the 
boundary state description indeed seizes some basic features of D-brane. In
fact the concept of the boundary state is introduced into string theory before 
that of D-brane. The BPS object can be represented by boundary state, which satisfies the
constraint $(Q+\tilde{Q})|B \rangle=0$. We can read off the tension and R-R
charge from the tree diagram of the boundary state. These are the most 
important proofs supporting the conjecture that  D-brane is the classical
soliton solution with R-R charge,  namely p-brane. Many relevant 
literatures have been included in \cite{long}. 

The concept of the boundary state is fundamental. Firstly, the tadpole can be
represented with the boundary state which is the source of closed string. 
Naively,  it seems reasonable that the boundary state should be
seen as quantum state of an independent dynamic object arising from  
string vacuum. D-brane is just defined as that object where open string 
end points end. Secondly, the boundary 
state can be defined according to the idea that one loop  amplitude of open string can
be regarded as tree amplitude of closed string. From this physical picture, 
the holographic principle \cite{witten} can be argued because
non-abelian gauge field theory appear from the low energy limits of open string theory 
and  the low energy limit of closed string theory can lead to  effective theory of gravity. 
Thirdly in \cite{cardy} Cardy generalizes the boundary state concept 
so as to get an exact
description of the boundary conformal field theory. Now the boundary conformal
field theory of  Gepner model, which is important in the 
heterotic string theory, has been proposed in \cite{reck}. This generalizes  D-brane 
concept, since the original D-brane only
comes from the system of  Type I string, Type IIA string and Type IIB string. 
Fourthly, in \cite{ishibashi}, the  path
integral representation of  boundary state provides a simple proof 
supporting the fundamental argument
in  Matrix theory that physics of $D_p$ brane is equivalent to that of  infinitely
many $D_{(p-2r)}$ branes. Fifthly, the path integral formalism has been incorporated 
into recent development of Noncommutative geometry in  string theory in 
\cite{ishibashii}.  Finally, in \cite{long} the classical $p$-brane solution can be constructed
from one point amplitude of  boundary state with closed string states.
This is also an important proof that  D-brane is the classical soliton solution
 with  R-R charge. 

Here we only give some examples of the path integral representation
of the conformal invariant boundary state~\footnote{
 The Feynman path integral is one way to represent a quantum theory, and it is 
 a very natural method for describing interactions in string theory.
Since we have had a systematic study of string theory using the Polyakov path integral,
we may try to explore the path integral representation of D-brane.}. 
Since the boundary conformal field theory is a candidate for D-brane theory, 
we must find  solutions of the boundary conformal field theory. In the case of
path integral representation, algebraic constraint equation can be transformed
into  differential equation which is easily solved to obtain the solution represented by string
modes. Although the 
conformal invariant boundary state satisfying Cardy condition  
can be constructed with Ishibashi state, this formalism is not often used in the practical
calculation. For example in those review articles \cite{long}
the mode representation of
the boundary state is used calculating various amplitudes~\footnote{ 
It is necessary to notice the path integral representation of boundary state 
is easily connected with  divergence.
Although the path integral representation of the boundary state in non-constant
$U(1)$ gauge field background in \cite{hash} can be obtained, 
those boundary states contain bad divergences in \cite{hash} and
so need regularization and renormalization. 
Another example is the path integral representation of
the boundary state in the linear dilaton background whose procedure of regularization 
and renormalization will be discussed in Appendix B. }. 

The organization of this paper is as follows.  In section 2, only from conformal invariance
we deduce the path integral representation of  boundary state in  constant $U(1)$ 
gauge field background~\footnote{ 
Although the path integral representation of boundary state 
in constant $U(1)$ gauge field background in \cite{clny} 
had been argued so as to determine the normalization factor, here the familiar 
form of  path integral representation in \cite{clny} can be deduced directly from  the conformal
invariance.}.
and show some applications of this formalism in T-duality and Matrix theory. 
In section 3, we construct the conformal invariant boundary state in the linear
dilaton background with the path integral representation, and at the same time construct
the path integral representation of the boundary state in open string tachyon background. 
Section 4 is devoted to  discussions.

\section{The path integral representation of the boundary state 
in the $U(1)$ gauge field background} 

The boundary state must be conformal invariant, which is the requirement of the 
reparametrization invariance on the world sheet.
So the conformal invariance can be directly used to construct the path integral 
representation of boundary state in the general background. 
In this section, firstly this procedure will be given in detail.
Secondly T-duality of D-brane to 
construct $D_p$ brane from  $D_{-1}$ brane can be carried out 
in the path integral formalism of the boundary state. 
Then the simplest solution of those differential equations is
found right to be the path integral representation of  boundary state in  
constant $U(1)$ gauge field background.

\subsection{The path integral representation of the boundary state}
In this paper only consider the bosonic string case. 
In fact the generalization to supersymmetrical case 
is straightforward. The holomorphic stress tensor reads
\begin{equation}
T=-{1\over 2}:(\partial {X})^2:, 
\end{equation}
and a similar anti-holomorphic counterpart.  
It is necessary to work with mode expansions 
\eqa
X(\tau, \sigma)&=&q+(\alpha_0+\tilde{\alpha}_0)\tau
 +i\sum_{n\ne 0}{1\over n}\left(\alpha_{n}e^{-in(\tau-\sigma)}+\tilde{\alpha}_{n}
e^{-in(\tau+\sigma)}\right), \\
X(z, \bar{z}) &=&q-i(\alpha_0 \hbox{ln}z+\tilde{\alpha}_0\hbox{ln}\bar{z})
 -i\sum_{n\ne 0}{1\over n}\left(\alpha_{-n}z^n+\tilde{\alpha}_{-n}
\bar{z}^n\right), \\
L_n &=&{1\over 2}\sum_{m}:\alpha_{m+n}\alpha_{-m}:
\ena
in which we have taken the Regge slope
$\alpha^{\prime}=2$. 
The commutators are $[\alpha_m, 
\alpha_n]=m\delta_{m+n, 0}$,  and similarly for the right-moving modes. 
Let $K_n=L_n-\tilde{L}_{-n}$.  The boundary condition is entirely
encoded in the boundary state $|B\rangle$,  and the conformal invariant
condition is $K_n|B\rangle=0$. 
To solve equations $K_n|B, p\rangle=0$, 
it  is essential to adopt the coherent state technique introduced
in \cite{clny}.  Introduce the following coherent state which satisfies
\begin{equation}
(\alpha_n-\tilde{\alpha}_{-n}-x_n)|x, p\rangle=0, \label{xp}
\end{equation}
where $n$ can be either positive or negative.  Further the conjugate coherent state
should satisfy
\begin{equation} 
\langle x, p|(\alpha_n-\tilde{\alpha}_{-n}-x_n)=0.  
\end{equation}
If requiring 
$\langle x, p|=(|x, p\rangle)^{\dag}$, then $x_{-n}=(x_{n})^{\dag}$. 
Solve this equation (\ref{xp}), the correspondent solution is
\begin{equation}
|x, p\rangle=\exp\left(\sum_{n=1}^\infty{1\over n}[-{1\over 2}
x_nx_{-n}+\alpha_{-n}\tilde{\alpha}_{-n}+x_n\alpha_{-n}-
x_{-n}\tilde{\alpha}_{-n}]\right)|p\rangle. 
\end{equation}
This set of states form a complete orthogonal basis as can be checked.
The formulas
$e^{A+B}=e^{A}e^{B}e^{{-1 \over 2}[A, B]}$, and $e^{A}e^{B}=e^{B}e^{A}e^{[A, B]}$
when $[[A, B], A]=[[A, B], B]=0$ 
are very helpful, and will be often used 
in the involved calculation in this paper. Observing this coherent state, we can 
find each oscillation mode can be traced in this state!  Basing on this point, 
our goal may be realized that transforming the algebraic equations into the 
differential equations. So the detailed operation is in essence a kind of substitution
\eqa
\alpha_{-n}|x, p \rangle =\left( n{\partial\over\partial{x_n}}+
 {1\over 2}x_{-n}\right)|x, p \rangle, 
\nonumber \\
\tilde{\alpha}_{-n}|x, p \rangle =\left(-n{\partial\over\partial{x_{-n}}}
-{1\over 2}x_{n}\right)|x, p \rangle,  \nonumber \\
\ena 
with the requirement of $ n \neq 0 $. 
We postulate that the boundary state with the boundary interaction
is of the form
\begin{equation}
|B, p\rangle=\int [dx]|x, p\rangle \exp{\left(S(x)\right)}, \label{B, p, S}
\end{equation}
in which $S(x)$ represents the special boundary interaction.
In fact the general form of the boundary interaction should admit 
string zero mode $\hat{q}$ or differential operator${\partial \over \partial{x}}$. 
 
To solve $K_n|B, p\rangle=0$,  one first computes
\eqa
K_n|x, p\rangle &=& \left( px_n +\sum_{m \neq {0, -n} }^{\infty}mx_{m+n}{\partial \over 
\partial x_m } \right) |x, p=\bar{p} \rangle, 
\ena
in which $n \neq 0$ is necessary. Substituting the above relations into $K_n|B, p\rangle=0$ 
and integrating
by parts,  the conformal invariance condition for $S(x)$ is transformed into
\begin{equation}
\left(-px_n+ 
   \sum_{m\neq {0, -n}}^\infty {m x_{m+n}{\partial\over\partial x_m }}\right) S(x)=0.  
   \label{diffn}
\end{equation}

However for $n=0$ case the thing is not similar.
Applying the above procedure to
\begin{equation}
K_0|B, p\rangle=0,
\end{equation}
we obtain the equation 
\begin{equation}
\sum_{m\neq 0}^\infty {m x_{m}{\partial\over\partial x_m }}S(x)=0. \label{diff0}
\end{equation}
To our special ansatz of the boundary interaction and the special choice of
 vacuum state,  the solution of the equation (\ref{diff0}) and the equation
(\ref{diffn}) corresponds to a special 
conformal invariant boundary state. In order to obtain one 
 boundary state, we may take various ansatz of  vacuum state
and  boundary interaction. For example, if we take  vacuum state
$|p=-\bar{p} \rangle $, then in the $n\neq 0$ case the equation is the form, 
\begin{equation}
\left( 
   \sum_{m\neq 0}^\infty {m x_{m+n}{\partial\over\partial x_m }}\right) S(x)=0.  
   \label{diffnn}
\end{equation}

\subsection{ T-duality of D-brane}
Since T-duality interchanges Neumann and Dirichlet boundary conditions, a further
T-duality in a direction tanget to a Dp-brane reduces it to a $D_{(p-1)}$ brane,
while a T-duality in an orthogonal direction turns it into a $D_{(p+1)}$ brane.
In the path integral representation of the boundary state of D-brane,
T-duality of $D_p$ brane can be  carried out.

The coherent state which satisfies
\begin{equation}
(\alpha_n-\tilde{\alpha}_{-n}-x_n)|x, p=-\bar{p}\rangle=0 \label{xp-p}
\end{equation}
has to satisfy the following constraint
\begin{equation}
(\alpha_n+\tilde{\alpha}_{-n})|x, p=-\bar{p}\rangle=
-2n{\partial\over\partial x_{-n}}|x, p=-\bar{p}\rangle. 
\end{equation} 
Therefore the boundary state 
\begin{equation}
|B~\rangle_N=\int [dx] |x, p=-\bar{p} \rangle 
\end{equation} 
satisfies Neumann condition 
$\partial_{\tau}X(\sigma)|B~\rangle_N=0$. 
By carrying out the path integral, its detailed formalism is  
\begin{equation}
|B~\rangle_N =A\exp\left(\sum_{n=1}^\infty{1\over n}[-\alpha_{-n}
  \tilde{\alpha}_{-n}]\right)|p=-\bar{p} \rangle, 
\end{equation} 
where A is a proportional constant. Further, another coherent state 
has to be also defined in order to realize our goal in this subsection.  
Firstly, some formulas should be given by
\eqa
X(\sigma)&=&q+i\sqrt{\alpha^{\prime}\over 2}\sum_{n\ne 0}{1\over n}
 \left(\alpha_{n}-\tilde{\alpha}_{-n}\right) e^{in\sigma},  \label{X}\\
\partial_{\tau}X(\sigma)&=&\alpha^{\prime} P+\sqrt{\alpha^{\prime}\over 2}\sum_{n\ne 0}
\left(\alpha_{n}+\tilde{\alpha}_{-n}\right) e^{in\sigma}, 
\ena
where $P$ is the total string momentum. 
Their commutator is $[X(\sigma), P(\sigma^{\prime})]=i\delta(\sigma-
\sigma^{\prime}) $,  in which the string momentum $P(\sigma)$ at the point labeled by the
$\sigma$ is defined as $
{1 \over 2\pi \alpha^{\prime} } \partial_{\tau}X(\sigma) $. The 
Fourier expansion of $\delta(\sigma-\sigma^{\prime})$ takes the form $
{1\over 2\pi}\sum_{m}\exp(-im(\sigma-\sigma^{\prime}))$. 
Now define a new
coherent state 
\begin{equation}
X(\sigma)|x\rangle=x(\sigma)|x\rangle.  \label{new}
\end{equation} 
The difference between the coherent state (\ref{xp}) and (\ref{new}) only lies in
that the latter is the position eigenstate $\hat{q} ~|x\rangle= q|x\rangle $
and the former is the momentum
eigenstate $\hat{P}|x, p\rangle= P|x, p\rangle $. 
Naturally, they may be connected by the transform 
\begin{equation}
\int Dx(\sigma) |x\rangle=\int [dx] |x, p=-\bar{p}\rangle. 
\end{equation}
And the coherent state (\ref{new}) can be constructed with the boundary state of
 instanton
namely  $D_{-1}$ brane,
\begin{equation}
|x\rangle=\exp\left(-i\int d\sigma P(\sigma)\cdot x(\sigma)\right) |B\rangle_{-1}, 
\label{ai}
\end{equation}
where the boundary state of instanton satisfies the constraint
$X(\sigma)|B\rangle_{-1}=0$. 
We conclude this subsection by constructing $D_p$ brane boundary state with 
$D_{-1}$ brane boundary state
\begin{equation}
|B\rangle_p=\int \prod_{i=0}^p Dx_i(\sigma) 
     \exp\left(-i\int d\sigma P^{i}(\sigma)\cdot x_{i}(\sigma)\right) 
            |B\rangle_{-1}  \label{fourier}
\end{equation}
with $i$ is limited to $ i=0,1, \cdot\cdot\cdot, p $~\footnote{
The idea used in \cite{ABB} where the path integral representation was used to support
open string T-duality has been generalized. Here as the result of a T-duality
in an orthogonal direction the boundary state of $D_p$ brane can be constructed from 
the boundary state of $D_{-1}$ brane. In (\ref{fourier})
a further T-duality in a direction tanget to a Dp-brane is equivalent to changing
the measure $\prod_{i=0}^p Dx_i(\sigma)$ to $\prod_{i=0}^{p-1} Dx_i(\sigma)$.
}. 
It can be checked
such  $D_p$ brane boundary state
will satisfy the boundary conditions, 
\eqa
&& X^{\mu}(\sigma)|B\rangle_p=0,  \mu=p+1, \cdot\cdot\cdot, D-1; \\
&& P^i(\sigma)|B\rangle_p=0,  i=0, 1, \cdot\cdot\cdot, p. 
\ena
On the other hand only from the above expression of (\ref{fourier}), 
T-duality  of 
boundary state seems to be Fourier transform in the configuration
space involved with T-duality. 

\subsection{The path integral representation of the boundary state in the 
$U(1)$ gauge field background}
We now turn to solve the differential equation (\ref{diff0}) and the differential
equation (\ref{diffnn}). It is easy to find one simple solution
\begin{equation}
S(x)={1\over 4}F_{\mu\nu}\sum_{m\neq 0}
  {{x_m}^{\mu}{x_{-m}}^{\nu} \over m}
\end{equation}
where $ x_{0}=0$ and $F_{\mu\nu}=-F_{\nu\mu} $ are necessary and  
the factor ${1\over 4}$ has been 
fixed due to the following calculation of the normalization factor. 
The correspondent boundary state is the form
\begin{equation}
|B, p=0\rangle=\int [dx] \exp\left({1\over 4}F_{\mu\nu}\sum_{m\neq 0}
  {{x_m}^{\mu}{x_{-m}}^{\nu} \over m}\right) |x, p=-\bar{p}=0\rangle. 
    \label{constant}
\end{equation}

Firstly this above solution~\footnote{
It is nontrivial to
point out that in the presence of a constant $U(1)$ gauge field background
Virasoro generators are not modified.}
 can be integrated out to show the final result is 
right the boundary state in the constant $U(1)$ gauge field background 
in \cite{long}.
The final result of (\ref{constant}) in the Euclidean spacetime
is
\begin{equation}
|B, p=0\rangle= N(F) \exp\left(-\sum_{n=1}^\infty{1\over n}{\alpha_{-n}
    {\left( 1-F \over 1+F \right)} \tilde{\alpha}_{-n}}\right)
    |p=-\bar{p}=0 \rangle, 
\end{equation} 
with the normalization factor $N(F)={\left( det(1+F) \right)}^{1 \over 2}$.
In bosonic string case 
the normalization factor is just the effective action of  gauge potential in\cite{tell}
~\footnote{We would like to thank Professor A.A.Tseytlin for bringing our attention
to the paper\cite{tell}. 
In fact, our purpose which is different from of \cite{tell} is to give an interesting 
example to show 
potential value of path integral representation. The familiar boundary state
in $U(1)$ constant gauge background can be only determined by conformal invariance.
The boundary state can be used to determine boundary condition from which the
action can be constructed. Naturally this example also shows the effective
action or the normalization factor can be determined directly from
the conformal invariance.}.

The purpose of arguing the path integral representation in \cite{clny}
is to obtain the normalization factor.
Here shows this approach to get normalization factor is rather natural in contrast
with the other methods. The method in \cite{long} is to compare the one point amplitude
of the boundary state with that from  Dirac-Born-Infeld action. In addition 
the detailed form of the normalization factor in \cite{hash} is in the requirement 
of gauge invariance of  D-brane source term contained in the action of  closed
string field theory. 

With the form (\ref{X}) and the following 
\begin{equation}
\partial_{\sigma}X(\sigma)=-\sum_{n\ne 0}
 \left(\alpha_{n}-\tilde{\alpha}_{-n}\right) e^{in\sigma},
\end{equation}
the path integral representation of (\ref{constant}) may be written with 
another form 
\begin{equation}
|B, p=0\rangle=\int [dx] \exp\left({i\over 8\pi}F_{\mu\nu}{\int_{0}}^{2\pi}
 d\sigma ~x^{\mu}(\sigma)\cdot\partial_{\sigma}x^{\nu}
     \right) |x, p=-\bar{p}=0\rangle
\end{equation}
which shows the boundary interaction of  $U(1)$ 
constant gauge field background. 

With the help of (\ref{ai}), the above boundary state can be changed into the form
\begin{equation}
|B, p=0\rangle =\int Dx(\sigma)
\exp\left({i\over 8\pi}F_{\mu\nu}{\int_0}^{2\pi}
 d\sigma ~x^{\mu}(\sigma)\cdot\partial_{\sigma}x^{\nu}
-i{\int_0}^{2\pi} d\sigma P_\mu\cdot x^{\mu}\right)|B\rangle_{-1} 
\end{equation}
from which the boundary condition determining such above 
boundary state may be found. 
Indeed,  one can show that the 
following identity holds:
\eqa
0
&=&
\int Dx(\sigma)\frac{\delta}{\delta x^{\mu}(\sigma )}
\exp\left({i\over 8\pi}F_{\mu\nu}{\int_{0}}^{2\pi}
 d\sigma ~x^{\mu}(\sigma)\cdot\partial_{\sigma}x^{\nu}
-i{\int_0}^{2\pi} d\sigma P_{\mu}\cdot x^{\mu}\right)|B\rangle_{-1}
\nonumber
\\
&=&
[{i\over 4\pi} F_{\mu\nu}\partial_{\sigma}X^{\nu}-iP_{\mu}(\sigma )]|B, p=0\rangle, 
\ena
where the boundary condition
\begin{equation}
\left(\partial_\tau X_{\mu}-F_{\mu\nu}\partial_\sigma X^{\nu} \right)|B, p=0\rangle=0  
\end{equation}
can be extracted.
Here it is both the boundary condition and the conformal invariance which require 
\begin{equation}
p=-\bar{p}=0. 
\end{equation}
Such the boundary condition just corresponds to the open string theory
\begin{equation}
S= \int  d\tau \int_0^{2\pi} d\sigma \left\{ \frac{1}{4\pi\alpha'}
[({\partial_{\tau} X})^2-({\partial_\sigma X})^2]-
[\delta(\sigma)-\delta(\sigma-2\pi)]{\dot X}^\mu A_\mu(X)\right\}. 
\end{equation}

The following is a simple review to \cite{ishibashi} where the path integral representation
of the boundary state in the $U(1)$ constant gauge field background can be also argued from
the point of Matrix theory.
Notice that their conventions are a little
different with ours since the definition of $X(\tau, \sigma)$ and  two order
antisymmetry tensor $F_{\mu\nu}$ are not completely fixed. 

The configuration of infinitely many D-instantons can be expressed by  
$\infty \times \infty$ hermitian matrices $X^M~(M=0, \cdots, D-1)$. 
The one we consider is 
\eqa
X^i
&=&
\hat{Q}^i, ~(i=0, \cdots, p)
\nonumber
\\
X^M
&=&
0~(M=p+1, \cdots, D-1), 
\label{PQ-1}
\ena
where $\hat{Q}^i~(i=0\cdots, p)$ satisfy 
\begin{equation}
[\hat{Q}^i, \hat{Q}^j]=i\theta^{ij}. 
\label{com}
\end{equation} 
Here  $(p+1)\times (p+1)$ matrix $\theta$ is assumed to 
be invertible.  

In  Matrix theory this configuration of D-instantons is equivalent to 
a D$p$-brane.  
A quick way to see the equivalence is to look at the boundary states.   
The boundary state $|B\rangle$ corresponding to the configuration 
eq. (\ref{PQ-1}) can be written as follows:
\begin{equation}
|B\rangle =\mbox{TrX}e^{-i\int_0^{2\pi}d\sigma P_i\cdot\hat{X}^i}|B\rangle_{-1}. 
\label{B}
\end{equation}

$|B\rangle_{-1}$ includes also  ghost part which is not relevant to the 
discussion here.  
The factor in front of $|B\rangle_{-1}$ is an analogue of  Wilson loop 
and corresponds to the background eq. (\ref{PQ-1}).  
Eq. (\ref{B}) can be rewritten with the path integral as
\begin{equation}
|B, p=0\rangle =\int Dx(\sigma)
\exp\left({\frac{i}{2}\int d\sigma~x^i\cdot\partial_\sigma x^j F_{ij}
-i\int d\sigma P_i\cdot x^i}\right)|B\rangle_{-1}, 
\label{path}
\end{equation}
where $F_{ij}=(\theta^{-1})_{ij}$.  

It is straightforward to perform the path integral in eq. (\ref{path}).  
In conclusion $|B\rangle$ is equivalent 
to the boundary state for a D$p$-brane with  $U(1)$ gauge field strength 
$F_{ij}$ on the worldvolume.  

\subsection{Some notes of the path integral representation of the boundary state }

To construct a good theory of $D_p$ brane in the general background,  we can argue
\begin{equation}
|B\rangle_p^N=\int \prod_{i=0}^p Dx_i(\sigma) 
     \exp\left(S(\hat{q}, x_n, {\partial\over \partial x_n}, \cdot\cdot\cdot)\right)
      |x(\sigma) \rangle,  \label{a111}
\end{equation}     
where  $|B\rangle_p^N$ is the Neumann part of the boundary state $|B\rangle_p$ which 
is defined as
\eqa
|B\rangle_p&=&|B\rangle_p^N\cdot|B\rangle_p^D
\ena
with $|B\rangle_p^D=\prod_{\mu=p+1}^{D-1} |B\rangle_{-1}^{\mu}$.
In essence (\ref{a111}) may be seen as Fourier transform between
the boundary state $|B\rangle_p^N$ with  boundary interaction and
the coherent state $|x(\sigma) \rangle$
~\footnote{From the above expression,  it can be argued
that physics of the world volume of $D_p$ brane, which is represented
by $|B\rangle_p^N$, may be 
equivalent to the complete physics of $(p+1)$ particles
in space time, which is represented by quantum state 
$\prod_{i=0}^{p} |x(\sigma) ~\rangle^{i}$.
The involved motion of these quantum states is with the weight 
$\left(S(\hat{q}, x_n, {\partial\over \partial x_n}, \cdot\cdot\cdot)\right)$
which is determined by the respective paths.}.

For the path integral representation of the boundary state
\begin{equation}
|B, p\rangle=\int [dx]|x, p=-\bar{p}\rangle \exp{\left(S(x)\right)}
\end{equation} 
we can check those equations  
\eqa
\left( 
   \sum_{m\neq 0}^\infty {m x_{m+n}{\partial\over\partial x_m }}\right) S(x)&=&0,~  
 n\neq0, \nonumber\\ 
\sum_{m\neq 0}^\infty {m x_{m}{\partial\over\partial x_m }}S(x)&=&0,  \nonumber\\ 
x_0&=&0
\ena
from the conformal invariance are consistent with the following equations 
\eqa
2n{\partial\over \partial x_n^{\mu}}S(x)&=&F_{\mu\nu}x_{-n}^{\nu},~n\neq 0, \nonumber\\
P&=&0 
\ena
from the mixed boundary condition determining the boundary state in the constant $U(1)$
gauge field background. 
Therefore when we take the ansatz $S(x)$,  we will know such path integral
representation
shows the physics of $D$ brane in constant $U(1)$ gauge
field background before the detailed path integral calculation in Appendix A. 
In fact it is important to firstly estimate the form of $S$ with
some physical consideration before determining the detailed form of $S$
by the conformal invariance.

In this paper some ansatz can be changed. Define the 
coherent state
\begin{equation} 
(\alpha_n+\tilde{\alpha}_{-n}-p_n)|{pp}\rangle=0  \label{xyz}
\end{equation}
then the boundary state with the boundary interaction can be expanded into the sum 
of the above coherent state, 
\begin{equation} 
|B, P\rangle=\int [dp]\exp
\left(S(\hat{q}, p_n, {\partial\over \partial p_n}, \cdot\cdot\cdot)\right)
  |{pp}\rangle=0.
\end{equation} 
So all the calculation in this paper can be carried out from the new starting point. 

\section{The path integral representation of the boundary state in other special backgrounds}

Besides the $U(1)$ gauge field background, the linear dilaton background and
the tachyon background are important in recent research. Especially, 
it seems attracting how
to construct the boundary state of D-brane in Type 0 string
under the closed string tachyon background
and make it consistent with the argument that D-brane is the classical soliton
solution with R-R charge. 

\subsection{The path integral representation of the boundary state in the linear 
dilaton background}
 We will construct the conformal invariant boundary state in the linear dilaton 
background with path integral representation
~\footnote{
Professor Miao Li introduced me his paper \cite{miao} and advised me to 
read this paper carefully. In \cite{miao} the conformal invariance was
directly used to construct the path integral representation of boundary
state in the linear dilaton background. However the path integral representation
of the boundary state in \cite{miao} can be found not to be exactly conformal
invariant. 
In the following the conformal invariant boundary state will be constructed.}.

Now the starting point is still the stress tensor
\begin{equation}
T=-{1\over 2}{(\partial\phi)^2+Q{\partial}^2\phi}\label{stress}
\end{equation}
and a similar anti-holomorphic counterpart. 
The central charge of
this free scalar is $c=1+12Q^2$,  and $Q=\sqrt{2}$ in two dimensional
string theory.  Consider a unit disk,  the conformal invariance condition
on the boundary means 
no net energy-momentum flow out of the boundary.
It is convenient to work with mode expansions
\begin{eqnarray}
\phi &=&\varphi_0-i(p\hbox{ln}z+\bar{p}\hbox{ln}\bar{z})
 -i\sum_{n\ne 0}{1\over n}\left(\alpha_{-n}z^n+\tilde{\alpha}_{-n}
\bar{z}^n\right), \cr
L_n &=&[\alpha_0+iQ(n+1)]\alpha_n+{1\over 2}\sum_{m\ne {0, -n}}\alpha_{m+n}\alpha_{-m}, 
~~~   n\neq{0}, \cr
L_0 &=&[{\alpha_0\over 2}+iQ]\alpha_0+{1\over 2}\sum_{m\ne 0}\alpha_{m}\alpha_{-m}.  
\end{eqnarray}
The similar formula for $\tilde{L}_n$.  The commutators are $[\alpha_m, 
\alpha_n]=m\delta_{m+n, 0}$,  and similarly for the right-moving modes. 
Let $K_n=L_n-\tilde{L}_{-n}$.  The boundary condition is entirely
encoded in the boundary state $|B\rangle$,  and the conformal invariance
condition is $K_n|B\rangle=0$. 

The usual Neumann boundary condition is given by $\partial_r\phi=0$
on the boundary of the unit disk.  In terms of the boundary state, 
it states that
$$P|B\rangle_N=(\alpha_n+\tilde{\alpha}_{-n})|B\rangle_N
=0. $$
Due to the existence of the background charge $Q$,  one has to modify
the boundary condition a bit: $p=-iQ$.  So there must be a net momentum
flow out of the boundary (in view of spacetime $\phi$).  One way to
see this is to consider the commutators
\begin{equation}
[K_m,  \alpha_n+\tilde{\alpha}_{-n}]=2iQm\delta_{m+n, 0}
-n\left(\alpha_{m+n}+\tilde{\alpha}_{-m-n}\right). 
\end{equation}
When the case $n+m=0$ occurs, the above commutator can be showed in the clearer
form
\begin{equation}
[K_m,  \alpha_n+\tilde{\alpha}_{-n}]=m(2iQ+\alpha_0+\tilde{\alpha}_0).  \label{ocomm}
\end{equation}
Actually, we have taken the following conventions,
\begin{equation}
2p=\alpha_0+\tilde{\alpha}_0, ~~~~~~\alpha_0=\tilde{\alpha}_0=p. 
\end{equation}
So when $p=-iQ$,  the center term disappears,  and it is possible to impose
both the conformal invariance condition and Neumann boundary
condition. To be as close to the ordinary Dirichlet condition as
possible,  one requires that a net momentum transfer is possible
if one scatters string states against the object described by
the boundary state.  So $|B, p\rangle$ is an eigen-state of $p$ with
arbitrary number $p$.  To solve equations $K_n|B, p\rangle=0$, 
 we may construct the path integral representation
of the conformal invariant boundary state.
 However the ansatz $S(x, \hat{q}, \hat{\tilde{q}})$ is
one little special form with the contribution of zero mode, 
\begin{equation}
S(x, \hat{q}, \hat{\tilde{q}})=
V_{coup}^{n}\exp\left({\hat{q}+\hat{\tilde{q}}\over 2Q}\right)\Phi(x), 
\end{equation}
where the operator $\hat{q}$ and $\hat{\tilde{q}}$ satisfy the commutative
relation $[\hat{q}, \hat{\tilde{q}}]=0$, $[\alpha_0, \hat{q}]=-i$ and $
[\tilde{\alpha}_0, \hat{\tilde{q}}]=-i$. And  $V_{coup}^{n}$ is the coupling 
constant before  renormalization procedure. 
The boundary state has been changed into the form 
\begin{equation}
|B \rangle=\int [dx] 
\exp\left(V_{coup}^{n}\exp\left({\hat{q}+\hat{\tilde{q}} 
  \over 2Q}\right)\Phi(x)\right)
   |x, p=\bar{p}=-iQ \rangle.  
\end{equation}
So the equation $K_n|B, p\rangle=0$ now can be changed into the following two
equations, 
\eqa
{-i \over 2Q} x_n\Phi(x)+2iQn^2{\partial \over \partial{x_{-n}} }\Phi(x)&=&  
\sum_{m\neq {0, -n}}^\infty {m x_{m+n}{\partial\over\partial x_m }} \Phi(x),  
 \\
\sum_{m\neq 0}^\infty {m x_{m}{\partial \over \partial x_m }}\Phi(x)&=&0. 
\ena
The special solution of the two equations could be found as follow, 
\begin{equation}
\Phi(x)=\oint {dz \over z}
  \exp\left(\sum_{m\neq 0} {i x_m  \over 2Qmz^{m}}\right). 
\end{equation}   
As we have claimed,
such above solution contains the divergence. Naturally the renormalized
boundary state which will be given in Appendix B is conformal invariant only
in some specific cases. 

Actually the conformal invariant boundary state in the linear dilaton
background can be constructed without considering the path integral
representation. Define the screening charge 
\begin{equation}
QQ=\oint dz :\exp\left(ik \cdot X(z)\right): \label{charge}
\end{equation}
with  $k$ must satisfy the following equation
\begin{equation}
k^2+2iQ\cdot k=2. 
\end{equation}
Since the commutative relation $[L_{n}, QQ]=0$, the conformal invariant
boundary state ~\footnote{This conformal invariant boundary state is useful in         
  Appendix B.    }
is 
\begin{equation}
|B \rangle=Fun(QQ) |B, p=\bar{p}=-iQ \rangle_{N} \label{dilaton}
\end{equation} 
in which $Fun$ is  function of $QQ$.  The boundary state 
$|B, p=\bar{p}=-iQ \rangle_{N}$ is the Neumanm boundary state with the momemtum which satisfies
\eqa
(\alpha_n+\tilde{\alpha}_{-n})|B, p=\bar{p}=-iQ \rangle_{N}&&=0,  n \neq 0,  \nonumber \\
(\alpha_0+\tilde{\alpha}_0)|B, p=\bar{p}=-iQ \rangle_{N}&&=2p|B, p=\bar{p}=-iQ \rangle_{N}
,  \nonumber  \\
(L_n-\tilde{L}_{-n})|B, p=\bar{p}=-iQ \rangle_{N}&&=0. 
\ena
Here the boundary state (\ref{dilaton})is still rather specifically.
In principle the number of  conformal invariant boundary state in the linear 
dilaton background is infinitely many.

\subsection{The path integral representation of the boundary state in the 
open string tachyon background}

The path integral representation of the boundary state in the 
open string tachyon background will be constructed, which provides some clues to
construct the conformal invariant boundary state of D-brane from Type $0$ string
in the tachyon background~\footnote{ It is necessary to point out that tachyon appearing
in Type 0 string theory is closed string state but here tachyon is open string state.
Since the boundary state is the closed string source, it is possible to construct
the boundary state of D-brane in Type 0 string.
On the other hand in order to support the argument the path
integral representation of boundary state can provide an exact description of 
D-brane in general background, we have to give a nontrivial example.
So the work to ensure the boundary state representing 
the boundary state of D-brane in Type 0 string is just nontrivial.
  }.

The conformal invariant boundary state in the open string tachyon background 
~\footnote{ The boundary state in the open string tachyon background has been researched 
    in \cite{callan1}. }
is the form
\begin{equation}
|B \rangle=\exp\left(\oint dz  :\exp\left(ik \cdot X(z)\right):\right)
  |B, p=-\bar{p}\rangle_N,   \label{tachyon}
\end{equation}
with the convention $k^2=2$. 
So the path integral representation of the boundary state 
in the open string tachyon background will have one form
\eqa
 |B \rangle=\int [dx]|x, p=-\bar{p}\rangle + \\
\int [dx]\sum_{l=1}^{\infty}\Phi_l|x, p=-\bar{p}\rangle. 
\ena
The differential equations from the conformal invariance constraint 
are
\begin{equation}
\Phi_l\left(-nl{\partial \over \partial x_{-n}}+{1\over 2}lx_{n}\right)
 +\Phi_l\left(\sum_{m \neq 0}mx_{m+n}
 {\partial \over \partial x_{m}}\right)=0, ~~n\neq 0,
\end{equation}
and for $n=0$ case the differential equation is
\begin{equation}
\Phi_l\left({1\over 2}{l^2}k^2+lk\cdot p\right)
 +\Phi_l\left(\sum_{m \neq 0}mx_{m}
 {\partial \over \partial x_{m}}\right)=0.
\end{equation}

Finally the ansatz of $\Phi_l$, namely
$\Phi_l\left(\hat{q}, p, x_n, x_{-n}, {\partial \over \partial x_n}\right) $ 
is solved as
\eqa
&& \Phi_l=\prod_{i=1}^{l}\oint dz_i\exp^{ik\cdot\hat{q}}z_i^{k\cdot p}z_{i}^{(i-1)k^2} 
\nonumber \\
&& \exp\left(\sum_{n=1}{{k\cdot x_n }\over{ -2nz_i^n}}\right)
 \exp\left(\sum_{n\ne 0}{k z_i^n\cdot{\partial \over {\partial x_n}}}\right)
  \exp\left(\sum_{n= 1}{{k\cdot x_{-n}z_i^n} \over  2n}\right),  \label{tttt}
\ena
with the definition of $\prod_{i=1}^{l}\oint [dz_{i}]\equiv
\oint [dz_1]\oint [dz_2]\cdot\cdot\cdot
\oint [dz_l]$.
In essence this solution is relevant with the form of (\ref{tachyon}).
The factor $z_{i}^{(i-1)k^2}$ appearing in the (\ref{tttt}) is   
from the noncommutative relation between  
zero modes, for example  $l=2$ case,
\eqa
&&\oint dz_2\exp^{ik\cdot\hat{q}}{z_2}^{k\cdot \hat{p}}
 \oint dz_{1}\exp^{ik\cdot\hat{q}}{z_1}^{k\cdot p}|x, p=-\bar{p}\rangle  \\
&& =\oint dz_1\exp^{ik\cdot\hat{q}}{z_1}^{k\cdot p}
 \oint dz_{2}\exp^{ik\cdot\hat{q}}{z_2}^{k\cdot p}{z_2}^{k^2}
 |x, p=-\bar{p}\rangle. 
\ena
Although such above calculation is not easy, it shows the path integral
representation of the boundary state is very useful if we can take some
tricks in the given background.

\section{Discussions}

In this paper some examples of the path integral representation of the
boundary state are given in some special backgrounds such as the $U(1)$ gauge field 
background, the linear dilaton background and 
the open string tachyon background. The path
integral representation of the boundary state in the $U(1)$ gauge field background
contains much essential information, especially the normalization factor which is
the effective action of  gauge field in the bosonic string theory. 
The application in Matrix theory and Noncommutative geometry hints that  Path
integral representation of the boundary state could  be rather fundamental in
describing  D-brane in the general background. 

The initial purpose of this paper is to construct a general solution of the 
boundary conformal field theory with  analytical approach,  mainly for the 
constraint equations $ (L_{n}-\tilde{L}_{-n}) |B \rangle =0 $ are difficult to
be solved to obtain the solution represented by  string 
modes from the pure algebraic approach. However in the path integral representation 
it is easy transforming those algebraic
equations into the differential formalism which can be solved.
In addition,  Cardy condition
which ensures the existence of  open string theory in the boundary conformal
field theory is
vital to this view point that the boundary conformal field theory is an exact
description of  D-brane in the general background. We will consider 
Cardy condition realization in the path integral representation in the future 
work. Finally, it is also our wish that the path integral representation of  
boundary state should be used to support or interpret the known ansatz in
\cite{witten}~\footnote{ It has been recently argued that holographic principle
should be deduced from noncommutative geometry, such as in \cite{chang}. And it seems
that two fundamental principles are intrinsic consistent in \cite{wu}. In fact they
can be both argued from the point of string/M theory. So it is possible to deduce
AdS/CFT duality from string/M theory. Naturally we wish that this procedure may be 
depend on the path integral representation of boundary state. From string amplitude with
D brane, which may be represented by saturating string state with boundary state
in\cite{long},
much useful information about supergravity can be obtained in\cite{kkk}. However those cases
are in flat spacetime, but now we have to face curved spacetime such as AdS.
We wish we could continue this research because this will provide a truely nontrivial
example of path integral representation of boundary state of D-brane.
},
\begin{equation}
\left\langle \exp\int_{S^d}\phi_0 O\right\rangle_{CFT}=Z_S(\phi_0).
\end{equation}

It is possible to supersymmetrize all the results in this paper. There
are still a lot of work  to be done. 
The path integral representation of  constant gauge 
field configuration should be generalized to  nonconstant background
in the conformal invariant
requirement.
The path integral representation of the boundary state in the linear
dilaton background might be useful to construct the boundary state under
 $AdS_{3}$ background, especially in the light cone gauge in \cite{yu}. 
Most important, the construction of the boundary state in  tachyon background
is potentially useful for the generalization of  $ AdS/CFT $ in  non-super
symmetrical case. The recent research on  D-brane in Type 0 string shows we
must face the tachyon problem in arriving at the purpose. However, it seems that the
usual boundary state is not reasonable which is constructed from the general
procedure without considering the tachyon background effect, because the classical
p-brane from the usual boundary state \cite{Divecchia} is not the solution 
in \cite{Klebanov}. This can
be the result of the introduction of the tachyon background, which makes the equations 
corresponding to  p-brane solution become nonlinear. But the usual boundary
state is only the linear combination of  closed string states. Since  
 nonsupersymmetrical generalization of $ AdS/CFT $ in  Type $0$ string case
seems more
reasonable than in that approach of supersymmetrical breaking case used in \cite{wittenn}, 
it is possible to construct the boundary state with the tachyon 
background in  Type $0$ string case.

\acknowledgments

We would like to thank Miao Li for helpful comments
on the manuscript and Yi-Hong Gao for a helpful discussion.

\bigskip
  
\noindent
\appendix
\section{ The calculation of the normalization factor $N(F)$ }
In the following the calculation of the normalization factor $N(F)$ is given.
The special solution of the differential equation (\ref{diff0}) and the differential
equation (\ref{diffnn}) is
\eqa
S(x)&=&{1\over 4}F_{\mu\nu}\sum_{m\neq 0}
  {{x_m}^{\mu}{x_{-m}}^{\nu} \over m},       \nonumber \\
&=&{-1\over 2}F_{\mu\nu}\sum_{m=1}^\infty
  {{x_{-m}}^{\mu}{x_{m}}^{\nu} \over m}.        
\ena
The entire calculation of the path integral of (\ref{constant}) has to be put  
in the Euclidean spacetime. The calculation of the (\ref{constant}) is as follow,
\eqa
|B, p=0\rangle&=&\int [dx] \exp\left({1\over 4}F_{\mu\nu}\sum_{m\neq 0}
 {{x_m}^{\mu}{x_{-m}}^{\nu} \over m}\right) |x, p=-\bar{p}=0\rangle, 
           \nonumber \\   \label{pathi}
&=&\int [dx] \exp\left({-1\over 2}\sum_{m=1}^\infty{1 \over m} (x_{-m}
-2\alpha_{-m}{1 \over 1+F})(1+F)(x_m
+2{1 \over 1+F}\tilde{\alpha}_{-m})\right)          \nonumber\\
& &\exp\left(-\sum_{m=1}^\infty{1\over m}{\alpha_{-m}
{\left( 1-F \over 1+F \right)} \tilde{\alpha}_{-m}} \right) |p=-\bar{p}=0\rangle. 
\ena
The integral measure in the above formula has to  be chosen as 
\eqa
\int [dx] &&\equiv \int \prod_{n=1}^\infty {dx_{-n}dx_n \over n},       \\
\int { dx_{-n}dx_n \over n} &&\equiv \int {dadb \over {(2\pi)}^{D}},            \label{dadb}
\ena
with the definition of $ x_n=\sqrt{n}(a+ib) $ and $ x_{-n}=\sqrt{n}(a-ib) $. 
The choice of (\ref{dadb}) aims at using the following formula in our
calculation, 
\begin{equation}
\int {dadb \over {(2\pi)}^{D}}\exp\left({-1\over 2}(a-ib)A(a+ib)\right)={1 \over A^D}.       
\end{equation}
So the final result can be given, 
\eqa
|B, p=0\rangle&=&\int [dx] \exp\left({1\over 4}F_{\mu\nu}\sum_{m\neq 0}
{{x_m}^{\mu}{x_{-m}}^{\nu} \over m}\right) |x, p=-\bar{p}=0\rangle,          \nonumber\\
&=&{\left(Det(1+F)\right)}^{-1}\exp\left(-\sum_{m=1}^\infty{1\over m}{\alpha_{-m}
{\left( 1-F \over 1+F \right)} \tilde{\alpha}_{-m}} \right)     \nonumber\\
& &  |p=-\bar{p}=0\rangle.                
\ena
In conclusion, the normalization factor should take the form
\begin{equation}
N(F)={\left(Det(1+F)\right)}^{-1}.          
\end{equation} 
Since 
\eqa
&&Det(1+F)=\prod_{m=1}^\infty~det(1+F), \nonumber\\
&&\sum_{m=1}^\infty {1}=\zeta(0)={-1 \over 2}, 
\ena
we obtain
\begin{equation}
N(F)={\left( det(1+F) \right)}^{1 \over 2}.
\end{equation}

\bigskip
  
\noindent

\section{ The conformal path integral representation of the boundary state in the linear dilaton
background  }
The path 
integral representation of the boundary state in the linear dilaton background 
is the form, 
\begin{equation}
|B \rangle=\int [dx] 
\exp\left(V_{coup}^{n}\exp\left({\hat{q}+\hat{\tilde{q}} 
  \over 2Q}\right)\Phi(x)\right)
   |x, p=\bar{p}=-iQ \rangle.  
\end{equation}
The conformal invariant constraint of the boundary state, 
$(L_n-\tilde{L}_{-n})|B \rangle=0$, can be expressed by
\eqa
(L_n-\tilde{L}_{-n})|B \rangle&=&\int [dx] 
\left[\alpha_0\alpha_n-\tilde{\alpha}_0\tilde{\alpha}_{-n}, \exp S(x, \hat{q}, \hat{\tilde{q}})\right]
 |x, p=\bar{p}=-iQ \rangle +    \nonumber \\
& & \int [dx] \exp S(x, \hat{q}, \hat{\tilde{q}}) \left( -2iQn^2{\partial \over \partial{x_{-n}}}+
\sum_{m\neq {0, -n}}^\infty {m x_{m+n}{\partial\over\partial x_m }} \right) \nonumber \\
& &|x, p=\bar{p}=-iQ \rangle=0.          
\ena
So the equation $K_n|B, p\rangle=0$ now can be changed into the following two
equations, 
\eqa
2iQn^2{\partial \over \partial{x_{-n}} }\Phi(x)&=&  
{i \over 2Q} x_n\Phi(x)+\sum_{m\neq {0, -n}}^\infty {m x_{m+n}{\partial\over\partial x_m }} \Phi(x),  
~~~n\neq 0, \\ \label{qqqq}
\sum_{m\neq 0}^\infty {m x_{m}{\partial \over \partial x_m }}\Phi(x)&=&0, 
\ena
with the commutative relation 
\eqa
&&\left[\alpha_0\alpha_n-\tilde{\alpha}_0\tilde{\alpha}_{-n}, \exp S(x, \hat{q}, \hat{\tilde{q}})\right] \nonumber\\
&&={-i \over 2Q}(\alpha_n-\tilde{\alpha}_{-n})S(x, \hat{q}, \hat{\tilde{q}})\exp S(x, \hat{q}, \hat{\tilde{q}}).       
\ena
The special solution of the two equations could be found, 
\begin{equation}
\Phi(x)=\oint {dz \over z}
  \exp\left(\sum_{m\neq 0} {i x_m  \over 2Qmz^{m}}\right).                
\end{equation}   
Now we come to check the solution. The right part of the (\ref{qqqq}) is
\eqa
&&{i \over 2Q} x_n\Phi(x)+
\sum_{m\neq {0, -n}}^\infty {m x_{m+n}{\partial\over\partial x_m }} \Phi(x)  \nonumber \\
&&=\oint dz~ (-z^n)~{ d \over dz}
  \exp\left(\sum_{m\neq 0} {i x_m  \over 2Qmz^{m}}\right);          
\ena
and the left part of the (\ref{qqqq}) is 
\eq
2iQn^2{\partial \over \partial{x_{-n}} }\Phi(x)=\oint dz~ {n z^{n-1} }
  \exp\left(\sum_{m\neq 0} {i x_m  \over 2Qmz^{m}}\right);      
\end{equation}
so the difference between the above expressions is  zero integral~\footnote{ 
if $x_0 \neq 0$ then the  conclusion
of the  zero integral is not correct.}. 
With 
$(\alpha_n-\tilde{\alpha}_{-n}-x_n)|x, p=\bar{p}=-iQ \rangle =0 $, 
and  $
\hat{q}|x, p=\bar{p}=-iQ \rangle =\hat{\tilde{q}}|x, p=\bar{p}=-iQ \rangle $,
 the boundary state is transformed into 
\eqa
|B \rangle&=&\int [dx] 
\exp\left(V_{coup}^{n}\exp\left({\hat{q}+\hat{\tilde{q}} 
  \over 2Q}\right)\Phi(x)\right)
   |x, p=\bar{p}=-iQ \rangle  ~~~~~~~\nonumber  \\
&=& \exp\left(V_{coup}^{n}\exp\left({\hat{q}+\hat{\tilde{q}} 
\over 2Q}\right)\oint {dz \over z}
\exp\left(\sum_{m\neq 0}{i (\alpha_m-\tilde{\alpha}_{-m}) \over 2Qmz^{m}}\right)\right) 
|B, p=\bar{p}=-iQ \rangle_{N} ~~~~~~~\nonumber  \\
&=& \exp\left(V_{coup}^{n}\exp({\hat{q} 
\over Q})\oint {dz \over z}
\exp\left(\sum_{m\neq 0}{i \alpha_m \over Qmz^{m}}\right)\right) 
|B, p=\bar{p}=-iQ \rangle_{N}          
\ena
which contains the divergence,  so needs normalizing and renormalization. 
With the commutator $[\alpha_n, \alpha_m]=n(1-\epsilon)^{|n|}\delta_{n+m, 0}$,  the expression
$\exp\left(\sum_{m\neq 0}{i \alpha_m \over Qmz^{m}}\right)$ may be represented by
\begin{equation}
\exp\left(\sum_{m=1}^\infty{i \alpha_{-m}z^{m} \over -Qm}\right)
\exp\left(\sum_{m=1}^\infty{i \alpha_{m} \over Qmz^{m} }\right)({\epsilon}^{{-1\over 2} Q^2} ).
\end{equation}
With the renormalized coupling constant $V_{coup}^{r}$,  the boundary state in the path
integral representation has the final form
\begin{equation}
|B \rangle=\exp\left(V_{coup}^{r}\exp({\hat{q} 
\over Q})\oint {dz \over z}
\exp\left(\sum_{m=1}^\infty{ \alpha_{-m}z^{m} \over iQm}\right)
\exp\left(\sum_{m=1}^\infty{ \alpha_{m} \over -iQmz^{m} }\right)\right) 
|B, p=\bar{p}=-iQ \rangle_{N}.          
\end{equation}
Since the boundary state $|B, p=\bar{p}=-iQ \rangle_{N}$ is conformal invariant, 
 we have to verify that
the former part of $|B, p=\bar{p}=-iQ \rangle_{N}$ should  commute
with Virasoro generator $L_n$ to ensure 
the conformal invariance of  $|B \rangle $ . 
In fact that part is similar to the screening charge (\ref{charge}), 
\eqa
QQ&=&\oint dz :\exp\left(ik \cdot X(z)\right): \nonumber \\
&=&\oint dz \exp({ik\cdot \hat{q}})~z^{k\cdot \hat{p}}
\exp\left(\sum_{m=1}^\infty{k\cdot\alpha_{-m}z^{m} \over m}\right)
\exp\left(\sum_{m=1}^\infty{k\cdot\alpha_{m} \over -mz^{m} }\right) \nonumber \\
& &|B, p=\bar{p}=-iQ \rangle_{N},             
\ena
with  $k$ must satisfy the following equation, 
\begin{equation}
k^2+2iQ\cdot k=2.       \label{kkk}
\end{equation}

With $k={1\over iQ }$, the screening charge is just the former factor of the 
first order term of  coupling constant of the boundary state.  In addition,  the 
solutions of  equation (\ref{kkk}) may be labeled by ${1 \over a_{+}}$ or 
${1 \over a_{-}}$. The $a_{+}$ and $a_{-}$ are respectively ${1\over 2}(Q+\sqrt{Q^2-2})$
and ${1\over 2}(Q-\sqrt{Q^2-2})$. So the 
solution $k=1/(iQ)$ is good in the large Q limit with the corrections 
suppressed by $1/Q^2$ etc. to $k$. 
Such a limit will change the 
boundary state$|B \rangle $  into $|B, p=\bar{p}=-iQ \rangle_{N}$ which 
is conformal invariant. Therefore it is reasonable 
that the renormalized boundary state should be conformal invariant in some 
special backgrounds.


\begin{thebibliography}{99}

\bibitem{pol}
J.  Dai,  R.  G.  Leigh and J.  Polchinski,  Mod.  Phys.  Lett.  {\bf A4} (1989) 2073;\\
J.  Polchinski,  hep-th/9510017,  Phys.  Rev.  Lett.  {\bf 75} (1995) 4724. 

\bibitem{long}
Matthias R Gaberdiel, hep-th/0005029;\\
Ben Craps, hep-th/0004198;\\
Paolo Di Vecchia,  Antonella Liccardo,  hep-th/9912161,  hep-th/9912275. 

\bibitem{witten}
Edward Witten, hep-th/9802150, Adv. Theor. Math. Phys.  2 (1998) 505-532. 

\bibitem{cardy}
John L.  Cardy,  Nucl.  Phys.  B324 (1989) 581.  \\
N. Ishibashi, 
Mod.  Phys.  Lett. A4 (1989) 251;  \\
N.  Ishibashi and T.  Onagi, 
Mod.  Phys.  Lett. A4 (1989) 161. 

\bibitem{reck}
A.  Recknagel,  V.  Schomerus, hep-th/9712186, 
Nucl. Phys.  B531 (1998) 185-225. 

\bibitem{ishibashi}
Nobuyuki Ishibashi,  hep-th/9804163, Nucl. Phys.  B539 (1999) 107-120.

\bibitem{ishibashii}
N.  Ishibashi,  H.  Kawai,  Y.  Kitazawa,  A.  Tsuchiya, hep-th/9908141, 
Nucl. Phys.  B565 (2000) 176-192; \\
N.  Ishibashi, hep-th/9909176; \\
Kazumi Okuyama, hep-th/9910138, JHEP 0003 (2000) 016;\\
Mitsuhiro Kato,  Tsunehide Kuroki, hep-th/9902004,  JHEP 9903:012, 1999. 

\bibitem{hash}
Koji Hashimoto,  Hiroyuki Hata, hep-th/9704125, 
Phys. Rev.  D56 (1997) 5179-5193;  \\
Koji Hashimoto,  hep-th/9909027, Phys. Rev.  D61 (2000) 106002;\\
Koji Hashimoto,  hep-th/9909095. 

\bibitem{clny}
C.  G.  Callan,  C.  Lovelace,  C.  R.  Nappi
and S.  A.  Yost,  Nucl.Phys.B 293 (1987) 83;
~~~Nucl.  Phys.  B 308 (1988) 221. 

\bibitem{ABB}
E.  Alvarez,  J.  L.  F.  Barbon,  J.  Borlaf, hep-th/9603089, 
Nucl. Phys.  B479 (1996) 218-242. 

\bibitem{tell}
E.S.Fradkin and A.A.Tseytlin, Phys.Lett.163B (1985) 123;\\
A.Abouelsaood, C.G.Callan, C.R.Nappi, and S.A.Yost, 
Nucl. Phys.B280 (1987) 599-624; \\
A. A. Tseytlin,  hep-th/9908105.

\bibitem{miao}
Miao.  Li,  hep-th/9512042,  
Phys. Rev.  {\bf D54 } (1996) 1644-1646. 

\bibitem{callan1}
Ali Yegulalp,  hep-th/9504104, Nucl.  Phys.  B450 (1995) 641;\\
G.  Callan,  Igor R.  Klebanov,  Andreas W.  W.  Ludwig and Juan M.  Maldacena, 
hep-th/9402113, Nucl.  Phys.  B422 (1994) 417-448. 

\bibitem{chang} 
Zhe Chang, hep-th/9904101, Phys.Rev. D61 (2000) 044009.

\bibitem{wu}
Miao Li, Yong-Shi Wu, Phys.Rev.Lett. 84 (2000) 2084-2087, hep-th/9909085.

\bibitem{kkk}
I.R. Klebanov and L. Thorlacius, Phys. Lett. {\bf B371} (1996)
51 [hep-th/9510200];\\
S.S. Gubser, A. Hashimoto, I.R. Klebanov and J.M. Maldacena,
Nucl. Phys. {\bf B472} (1996) 231 [hep-th/9601057]. \\
M.R. Garousi and R.C. Myers,
{ Nucl. Phys.} {\bf B475} (1996) 193 [hep-th/9603194]. \\
A. Hashimoto and I.R. Klebanov, 
{Phys. Lett.} {\bf B381} (1996) 437 [hep-th/9604065].\\
For a review, see: A. Hashimoto and I.R. Klebanov, Nucl. Phys. Proc.
Suppl. {\bf 55B} (1997) 118 [hep-th/9611214].\\
M.R. Garousi and R.C. Myers, ``World-Volume Interaction on D-branes,''
hep-th/9809100. \\
M.R. Garousi,``Superstring Scattering from D-brane Bound States,''
hep-th/9805078.

\bibitem{yu}
Ming Yu,  Bo Zhang, hep-th/9812216,  Nucl. Phys.  B551 (1999) 425-449. 

\bibitem{Divecchia}
P.  Di Vecchia,  M.  Frau,  I.  Pesando,  hep-th/9707068, 
Nucl. Phys.  B507 (1997) 259-276. 

\bibitem{Klebanov}
I.  R.  Klebanov,  A.  A.  Tseytlin, hep-th/9811035, 
Nucl. Phys.  B546 (1999) 155-181.   \\
Joseph A.  Minahan, hep-th/9902074, JHEP 9904 (1999) 007. 

\bibitem{wittenn}
Edward Witten,  hep-th/9803131, Adv. Theor. Math. Phys.  2 (1998) 505-532. 



\end {thebibliography}
\end{document}